\documentclass[12pt,preprintnumbers,superscriptaddress,amsmath,amssymb,nofootinbib]{revtex4}

\usepackage{graphicx}
\usepackage{dcolumn}
\usepackage{bm}
\usepackage{amssymb}
\usepackage{amsmath}
\usepackage{epsfig}    
\usepackage{color}
\usepackage{slashed}
\usepackage[hypertex]{hyperref}
\begin{document} 

\title{Higgs boson couplings in multi-doublet models with\\ natural flavour conservation}

\author{Kei Yagyu}
\affiliation{School of Physics and Astronomy, University of Southampton, Southampton, SO17 1BJ, United Kingdom}
\email{K.Yagyu@soton.ac.uk}

\begin{abstract}

We investigate the deviation in the couplings of the standard model (SM) like Higgs boson ($h$) with a mass of 125 GeV from the 
prediction of the SM in multi-doublet models within 
the framework where flavour changing neutral currents at the tree level are naturally forbidden. 
After we present the general expressions for the modified gauge and Yukawa couplings for $h$, 
we show the correlation between the deviation in the Yukawa coupling 
for the tau lepton $h\tau^+\tau^-$ and that for the bottom quark $hb\bar{b}$ under the assumption of a non-zero deviation in the $hVV$ $(V=W,Z)$ couplings
in two Higgs doublet models (2HDMs) and three Higgs doublet models (3HDMs) as simple examples.  
We clarify the possible allowed prediction of the deviations
in the 3HDMs which cannot be explained in the 2HDMs even taking into account the one-loop electroweak corrections to the Yukawa coupling. 

\end{abstract}
\maketitle

\section{Introduction}

The observation of the discovered Higgs boson $h(125)$ 
with decaying into a pair of tau leptons~\cite{ATLAS_tau,CMS_tau} provided  
more evidence for the existence of at least one $SU(2)_L$ doublet scalar field in addition to various other measurements of the properties of $h(125)$ 
at the CERN Large Hadron Collider (LHC). 
Then, the natural question is ``how many doublet fields are there in the Higgs sector?'' 
In general, multi-Higgs doublet models (MHDMs) can reproduce the predictions of the standard model (SM) composed of only one Higgs doublet, and so
we should take into account the possibility of the existence of multi-doublets. 

On the other hand, multi-doublet fields arise in many models beyond the SM. 
As the most familiar example, in the minimal supersymmetric SM (MSSM) 
the Higgs sector is extended to have two doublets because of the requirement of gauge anomaly cancellation. 
In addition, the multi-doublet structure is required to have an additional source of CP-violation, which is necessary to realize the successful scenario of  
electroweak baryogenesis~\cite{EWBG1,EWBG2,EWBG3}. 
Moreover, additional doublets are often introduced in various loop-induced neutrino mass models, such as the model by Zee~\cite{Zee} and by Ma~\cite{Ma} etc. 
The point is that the structure of MHDMs depends on new physics scenarios, e.g., in the MSSM the so-called Type-II Yukawa interaction is adopted. 
Therefore, by studying the phenomenology of MHDMs with a bottom-up approach, we can narrow down the possible scenarios of new physics beyond the SM. 

One of the most important issues when MHDMs are discussed is the possibility of 
flavour changing neutral currents (FCNCs) mediated by neutral Higgs bosons, which appear at the tree level in general. 
The simplest way to avoid such FCNCs is to realize a Yukawa Lagrangian where 
each type of fermion, i.e., charged leptons, up-type and down-type quarks  
couples to only one scalar doublet field. 
This is usually achieved by imposing discrete symmetries in the Higgs sector, and is so-called ``natural flavour conservation'' (NFC)~\cite{GW}. 
Depending on how doublet fields couple to each type of fermions, there are four (five) independent types of Yukawa interactions in models with 
two (more than two) doublets~\cite{Grossman}. 

In this paper, we discuss the deviation in the couplings of $h(125)$ to weak bosons ($hVV$ and $V=W,Z$) and fermions $(hf\bar{f})$ from the SM prediction 
in the MHDMs with NFC. 
In Ref.~\cite{Finger}, it has been clarified that the four types of Yukawa interactions in two Higgs doublet models (2HDMs) 
can be determined by measuring the correlation between the deviations in the charged lepton Yukawa couplings $he\bar{e}$ and the 
down-type quark Yukawa couplings $hd\bar{d}$ as long as there is a non-zero deviation in the $hVV$ couplings. 
Now, we extend this discussion to models with $N$ Higgs doublet fields, and we investigate how the pattern of the deviation can be different in 2HDMs and in MHDMs with $N\geq 3$. 

This paper is organized as follow. 
In Sec.~II, we define the MHDMs with NFC. 
We then give the general expressions for the $hVV$ and $hf\bar{f}$ couplings. 
We show more explicit forms of these couplings in the 2HDMs and the three Higgs doublet models (3HDMs) as simple examples. 
In Sec.~III, we give numerical results of the deviation in the couplings of $h(125)$ especially for the $h\tau^+\tau^-$ and $hb\bar{b}$ couplings in the 2HDMs and the 3HDMs. 
Conclusions are given in Sec.~IV.

\section{Multi-Doublet Models}

\subsection{Higgs Basis}

We consider models with $N$ Higgs doublet fields\footnote{The number of doublets can be constrained from the electroweak oblique $S$ and $T$ parameters. 
In Ref.~\cite{Hernandez}, the maximally allowed number of doublets has been given as a function of a mass difference among Higgs bosons. } $\Phi_i$ $(i=1,\dots,N)$ with hypercharge\footnote{The electric charge $Q$ is given by $Q=T_3+Y$ with $T_3$ being 
the third component of the isospin. } $Y=+1/2$ and vacuum expectation values (VEVs)
$v_i\equiv \sqrt{2}\langle \Phi_i^0\rangle$ which are taken to be real. 
Throughout this paper, we assume that the Higgs sector has neither explicit nor spontaneous CP-violation for simplicity. 
In order to extract couplings among physical Higgs bosons and gauge bosons or fermions, 
it is convenient to define the so-called Higgs basis~\cite{HB}, where 
only one of the $N$ doublets has the VEV $v$ which is related to the Fermi constant $G_F$ by 
$v = (\sqrt{2}G_F)^{-1/2}\simeq 246$ GeV. 
The Higgs basis is defined via the $N\times N$ orthogonal matrix $R$ by 
\begin{align}
\begin{pmatrix}
\Phi_1 \\
\Phi_2 \\
\vdots \\
\Phi_N
\end{pmatrix} = R
\begin{pmatrix}
\Phi \\
\Psi_2 \\
\vdots \\
\Psi_{N}
\end{pmatrix}.  \label{HB}
\end{align}
The doublets $\Phi$ and $\Psi_a$ $(a=2,\dots,N)$ appearing in the right-hand side of Eq.~(\ref{HB}) 
are expressed as
\begin{align}
\Phi = \begin{pmatrix}
G^+ \\
\frac{\tilde{H} +v + i G^0 }{\sqrt{2}}
\end{pmatrix},\quad 
\Psi_a = \begin{pmatrix}
\tilde{H}_a^+ \\
\frac{\tilde{H}_a + i \tilde{A}_a }{\sqrt{2}}
\end{pmatrix},  
\end{align}
where $G^\pm$ and $G^0$ are the Nambu-Goldstone bosons which are absorbed into the longitudinal components of the $W^\pm$ and $Z$ bosons, respectively. 
In Eq.~(\ref{HB}), the matrix $R$ is expressed in terms of $N-1$ angles, so that 
the $N$ VEVs are translated into $v$ and the $N-1$ angles. By using $R$, each of VEVs is expressed  as 
\begin{align}
v_i = R_{i1}v, 
\end{align}
and the sum rule: $\sum_i v_i^2 = v^2\sum_i R_{i1}^2 = v^2$ follows from this equation. 
In the Higgs basis, 
the $N-1$ pairs of singly-charged states $\tilde{H}_a^\pm$, $N-1$ CP-odd states $\tilde{A}_a$ and 
$N$ CP-even states $\tilde{H}_i \equiv  (\tilde{H},\tilde{H}_a)$ are not mass eigenstates in general. 
Their mass eigenstates can be defined by introducing $(N-1) \times (N-1)$ unitarity matrices $R_\pm$ and $R_A$
for the singly-charged states and CP-odd states and an $N\times N$ unitary matrix $R_H$ for the CP-even states as 
\begin{align}
\tilde{H}^\pm_a = (R_\pm)_{ab} H_b^\pm, \quad 
\tilde{A}_a = (R_A)_{ab}A_b, \quad
\tilde{H}_i = (R_H)_{ij}H_j. 
\end{align}
We identify the $h\equiv H_1$ state as the discovered Higgs boson with a mass of 125 GeV. 

Let us move on to the construction of the kinetic term and the Yukawa Lagrangian for the scalar doublets in the Higgs basis. 
Without loss of generality, the kinetic term is expressed as follows:
\begin{align}
{\cal L}_{\text{kin}} = |D_\mu \Phi|^2 + \sum_{a=2}^{N}|D_\mu \Psi_a|^2,  \label{kin}
\end{align}
where $D_\mu$ is the covariant derivative for the doublet Higgs fields. 
From the above Lagrangian, the gauge-gauge-scalar type interaction is obtained from the first term, because
only $\Phi$ contains the VEV $v$. 
Thus, we can extract the couplings of $h$ with weak bosons as 
\begin{align}
g_{hVV} = (R_H)_{11}\times g_{hVV}^{\text{SM}},~~V = W,~Z,  \label{ghvv}
\end{align}
where we take $0\leq (R_H)_{11}\leq 1$. 
We note that from the second term in Eq.~(\ref{kin}), 
we obtain the scalar-scalar-gauge type interactions which are proportional to the derivative of a scalar field.  
These interaction terms can be important when we consider production and decay of extra Higgs bosons. 

Next, we give the Yukawa Lagrangian. 
In the scenario based on NFC, the Yukawa Lagrangian takes the following form:
\begin{align}
-{\cal L}_Y =  \bar{Q}_L \,Y_u \, \tilde{\Phi}_u u_R^{} 
+ \bar{Q}_L  \,Y_d \, \Phi_d d_R^{}
+ \bar{L}_L  \,Y_e \,\Phi_e e_R^{} + \text{h.c.},  \label{yuk}
\end{align}
where $\Phi_{u,d,e}$ are any one of $\Phi_i$, and $\tilde{\Phi}=i\sigma_2\Phi^*$ with $\sigma_2$ being the second Pauli matrix. 
Here, we do not explicitly show the flavour indices. 
The Lagrangian given in Eq.~(\ref{yuk}) is naturally realized by imposing discrete symmetries such as $Z_N$ for models with $N$ Higgs doublet fields. 
Under such a discrete symmetry, all the $N$ doublets can have a distinct charge from each other. 
Depending on the charge assignment for the right-handed fermions\footnote{The charge assignment for the right-handed fermions 
should be taken as the flavour blind way to obtain the Lagrangian given in Eq.~(\ref{yuk}). }, there appear five independent types of Yukawa interactions. 
They can be defined as follows:
\begin{subequations}
\begin{align}
&\text{Type-I}:  \Phi_u = \Phi_d = \Phi_e, \\
&\text{Type-II}: \Phi_u \neq \Phi_d,\quad \Phi_d = \Phi_e, \\
&\text{Type-X}:  \Phi_u = \Phi_d, \quad  \Phi_d  \neq \Phi_e, \\
&\text{Type-Y}:  \Phi_u \neq \Phi_d,\quad  \Phi_u = \Phi_e, \\
&\text{Type-Z}:  \Phi_u \neq \Phi_d,\quad \Phi_d \neq \Phi_e,  \quad \Phi_e \neq \Phi_u. 
\end{align}\label{types}
\end{subequations}
A similar classification of the Yukawa interactions has been presented in Ref.~\cite{Ferreira}. 
We note that the Type-II, Type-X and Type-Y (Type-Z) are realized for $N\geq 2$ ($N\geq 3$).

The Yukawa Lagrangian can be rewritten in the Higgs basis as
\begin{align}
-{\cal L}_Y = 
&\frac{\sqrt{2}}{v}\Bigg[\bar{Q}_L M_u\left(\tilde{\Phi} + \sum_{a=2}^N\xi_u^a \tilde{\Psi}_a \right) u_R^{} 
+ \bar{Q}_L M_d\left(\Phi + \sum_{a=2}^N\xi_d^a \Psi_a \right) d_R^{} \notag\\
&+ \bar{L}_L M_e\left( \Phi + \sum_{a=2}^N\xi_e^a \Psi_a \right) e_R^{} \Bigg]
+ \text{h.c.}, 
\end{align}
where $M_u$, $M_d$ and $M_e$ are respectively the mass matrix for up-type quarks, down-type quarks and charged leptons, which 
are given by $M_F = vY_FR_{F1}/\sqrt{2}$ for $F=u,d,e$. 
The $\xi_F^a$ factors are determined by the matrix elements of $R$ as follows:
\begin{align}
\xi_F^a = \frac{R_{Fa}}{R_{F1}},~~a=2,\dots, N. \label{xif}
\end{align}
In terms of $(R_H)_{ij}$ and $\xi_F^a$, the Yukawa couplings for $h$ are expressed by 
\begin{align}
y_{hFF} =  \left[(R_H)_{11} + \sum_{a = 2}^N(R_H)_{a1}\xi_F^a \right] \times y_{hFF}^{\text{SM}}. \label{yhff}
\end{align}

In order to express the deviations in the Higgs boson couplings, we introduce the scaling factor for the gauge couplings by 
$\kappa_V^{} \equiv g_{hVV}^{}/g_{hVV}^{\text{SM}}$ and for the Yukawa couplings by $\kappa_F^{} \equiv y_{hFF}^{}/y_{hFF}^{\text{SM}}$. 
We can express $\kappa_F^{}$ using $\kappa_V=(R_H)_{11}$ by 
\begin{align}
\kappa_F^{} = \kappa_V^{} +   \sum_{a = 2}^N(R_H)_{a1}\xi_F^a ,~~\text{with}~~ \sum_{a = 2}^N(R_H)_{a1}^2 = 1 - \kappa_V^2.  \label{kappaf}
\end{align}
From the above expression, it is seen that $\kappa_F^{}$ become unity by taking the limit of $\kappa_V^{}\to 1$, and 
this limit is the so-called ``alignment limit''~\cite{alignment-limit}. 
We thus can regard $h$ as the SM-like Higgs boson for the case with $\kappa_V^{}\simeq 1$. 
Because of the existence of the alignment limit, the MHDMs can safely reproduce predictions in the SM. 
Notice here that in MHDMs, $\kappa_V^{}$ must be smaller or equal to unity at the tree level\footnote{If 
we introduce higher $SU(2)_L$ multiplets such as triplets, $\kappa_V^{}>1$ is possible at the tree level~\cite{kv1}. }. On the other hand, 
$|\kappa_F^{}|$ can be both smaller and larger than unity due to the second term of Eq.~(\ref{kappaf}). 

\subsection{Examples}

\begin{table}[t]
\begin{center}
\begin{tabular}{c||ccc||ccc|ccc}\hline\hline
&\multicolumn{3}{c||}{2HDM}&\multicolumn{6}{c}{3HDM}\\\hline
          & $\xi_u$     & $\xi_d$      & $\xi_e$      & $\xi_u^2$ & $\xi_d^2$ & $\xi_e^2$ & $~~\xi_u^3~~$ & $\xi_d^3$ & $\xi_e^3$          \\  \hline
Type-I    & $\cot\beta$ & $\cot\beta$  & $\cot\beta$ & $\cot\beta_1$&$\cot\beta_1$&$\cot\beta_1$&0&0&0  \\  \hline
Type-II     &$\cot\beta$&$-\tan\beta$&$-\tan\beta$&$\cot\beta_1$&$-\tan\beta_1$&$-\tan\beta_1$&0&$-\tan\beta_2/\cos\beta_1$&$-\tan\beta_2/\cos\beta_1$  \\  \hline
Type-X      &$\cot\beta$&$\cot\beta$ &$-\tan\beta$&$\cot\beta_1$&$\cot\beta_1$ &$-\tan\beta_1$&0&0&$-\tan\beta_2/\cos\beta_1$ \\  \hline
Type-Y      &$\cot\beta$&$-\tan\beta$ &$\cot\beta$&$\cot\beta_1$&$-\tan\beta_1$ &$\cot\beta_1$&0&$-\tan\beta_2/\cos\beta_1$&0  \\  \hline
Type-Z      &--&-- &--&$\cot\beta_1$&$-\tan\beta_1$  &$-\tan\beta_1$&0&$-\tan\beta_2/\cos\beta_1$&$\cot\beta_2/\cos\beta_1$  \\  \hline\hline
\end{tabular} 
\end{center}
\caption{$\xi_F^a$ factors appearing in Eq.~(\ref{yhff}) for each type of Yukawa interaction in the 2HDM and the 3HDM. 
In the 2HDM, the $\xi_F^2$ factors are rewritten by $\xi_F$. }
\label{ratios}
\end{table}

Based on the general discussion given in the previous subsection, we here 
consider two simple examples, i.e., the case for $N=2$ (2HDMs) and for $N=3$ (3HDMs). 

In the 2HDMs\footnote{For a comprehensive review on 2HDMs, see Ref.~\cite{Branco}.  }, the matrix $R$ takes the $2\times 2$ form as 
\begin{align}
R = \begin{pmatrix}
c_\beta & -s_\beta \\
s_\beta & c_\beta
\end{pmatrix},~~\text{for}~~N=2, \label{beta}
\end{align}
where $\tan\beta \equiv v_2/v_1$. We introduce shorthand notations for the trigonometric functions as $c_\theta^{}$ ($s_\theta^{}$)  $=\cos \theta~ (\sin\theta)$. 
From Eqs.~(\ref{xif}) and (\ref{beta}), we can find that $\xi_F^2$ (notice that the superscript 2 does not mean the square of $\xi_F^{}$) is taken to be either $\cot\beta$ or $-\tan\beta$ depending on the type of Yukawa interaction and the type of fermion. 
If we fix $\Phi_u = \Phi_2$, all the $\xi_F^{} \equiv \xi_F^2$ factors are fixed for each type of Yukawa interaction and for each type of fermion as shown in Table~\ref{ratios}.
In addition, the matrix $R_H$ also takes $2\times 2$ form with an angle\footnote{This angle depends on parameters in the scalar potential. 
In this paper, we do not explicitly show the potential whose structure depends not only on the number of doublets but also on the symmetry to avoid the tree level FCNCs.} independent of 
$\beta$. 
Conventionally, this angle is expressed by $\beta-\alpha$ (see e.g., \cite{HHG}), so that $R_H$ is written as 
\begin{align}
R_H = \begin{pmatrix}
s_{\beta-\alpha} & c_{\beta-\alpha} \\
c_{\beta-\alpha} & -s_{\beta-\alpha} 
\end{pmatrix}. 
\end{align}
Because of the orthogonal property of the $R_H$ matrix, 
$(R_H)_{21} (= c_{\beta-\alpha} )$ appearing in Eq.~(\ref{kappaf}) is simply written by $\sigma \sqrt{1-\kappa_V^2}$ with $\sigma = \text{Sign}[(R_{H})_{21}]$, 
and thus $\kappa_F^{}$ is determined by $\tan\beta$, $\kappa_V^{}$ and $\sigma$. 
Then, $\kappa_F^{}$ can be rewritten by 
\begin{align}
\Delta \kappa_F^{} = -\Delta\kappa_V^{} + \sigma \,\xi_F \sqrt{(2-\Delta\kappa_V^{})\Delta\kappa_V^{}} 
\simeq  -\Delta\kappa_V^{} +\sigma \, \xi_F\sqrt{2\Delta \kappa_V^{}} , \label{kappaf2}
\end{align}
where $\Delta\kappa_{V}^{}\equiv 1- \kappa_V^{}~(\geq 0)$ and $\Delta \kappa_F \equiv \kappa_{F}^{} -1$. 
The far-right hand side is valid for $\kappa_V^{} \simeq 1$ which is supported by the current measurement at the LHC~\cite{LHC-combine}. 
We note that because of the $\sigma\,\xi_F$ term of Eq.~\ref{kappaf2}, 
we can take the so-called ``wrong sign limit''~\cite{wsl} defined by $\Delta \kappa_F \to -2$. 
The case with this limit or this regime provides a phenomenologically interesting scenario, where loop induced decay rates such as $h \to gg$
and $h\to \gamma\gamma$ can be modified from the SM prediction due to effects of interference even having the same or similar value of the decay rate of $h\to VV$ and $h\to f\bar{f}$
as those in the SM at the tree level. 

Eq.~(\ref{kappaf2}) tells us the following important fact. 
As long as $|\xi_F|>  \sqrt{\Delta\kappa_V^{}/(2-\Delta\kappa_V^{})}\,(\leq 1)$,  
the sign of $\Delta\kappa_F^{}$ is determined by the sign of $\sigma\,\xi_F^{}$. 
In this case, the prediction of $(\Delta\kappa_e,\Delta\kappa_d)$ in 
the four types of Yukawa interactions appears in the four different quadrant on the $\Delta\kappa_e$--$\Delta\kappa_d$ plane for a fixed $\sigma$ which can be 
determined by measuring the sign of $\Delta\kappa_u$. 
Namely for $\Delta\kappa_u >0$, the sign of $(\Delta\kappa_e,\Delta\kappa_d)$ is predicted by 
$(+,+)$, $(-,-)$, $(-,+)$ and $(+,-)$ in the Type-I, -II, -X and -Y Yukawa interaction, respectively, while for $\Delta\kappa_u < 0$
these signs are flipped as compared to the former case.
Therefore, by the precise measurements of $\Delta\kappa_e$ and $\Delta\kappa_d$, we can determine the type of Yukawa interaction in the 2HDM.

Although this statement is valid for the case with $|\xi_F|>  \sqrt{\Delta\kappa_V^{}/(2-\Delta\kappa_V^{})}$ for all $F$ as mentioned above, 
the case with at least one of $|\xi_F^{}|$ being smaller than $\sqrt{\Delta\kappa_V^{}/(2-\Delta\kappa_V^{})}$
is phenomenologically quite difficult to realize in the Type-II, -X and -Y 2HDM, which is explained following. 
In order to achieve\footnote{For $\xi_F= -\tan\beta$, the condition $|\xi_F| >  \sqrt{\Delta\kappa_V^{}/(2-\Delta\kappa_V^{})}$ 
is satisfied unless we take $\tan\beta < 1$ which is disfavored by the various $B$ physics constraints mainly due to the enhancement of the top Yukawa coupling. 
See, e.g., \cite{Stal,Watanabe} for the constraints on the parameter space from $B$ physics in 2HDMs. } 
$\xi_u (= \cot\beta) < \sqrt{\Delta\kappa_V^{}/(2-\Delta\kappa_V^{})}$, 
we need a rather large value of $\tan\beta$ if $\Delta \kappa_V \ll 1$. 
For example, when $\Delta\kappa_V^{}$ is given to be $5\% \, (1\%)$, 
$\tan\beta$ should be larger than about 6\, (14) to satisfy $\cot\beta < \sqrt{\Delta\kappa_V^{}/(2-\Delta\kappa_V^{})}$. 
The important point is that such a case gives a huge deviation in some of Yukawa couplings in the Type-II, -X, and -Y 2HDM. 
In fact in the Type-II 2HDM, $\Delta\kappa_e(=\Delta\kappa_d)\simeq 180\%\,(200\%)$ for the case of $\Delta \kappa_V^{}=5\%\,(1\%)$, 
$\tan\beta=6\,(14)$ and $\sigma = -1$. 
Similarly in the Type-X (Type-Y) 2HDM, $\Delta\kappa_e$ $(\Delta\kappa_d)$ becomes the above value, and 
it goes without saying that such a huge deviation has already been excluded by the current data at the LHC~\cite{LHC-combine}. 
Therefore, if $|\xi_F| <  \sqrt{\Delta\kappa_V^{}/(2-\Delta\kappa_V^{})}$ is realized, 
only the Type-I 2HDM provides a phenomenologically acceptable scenario.

Next, we consider the 3HDMs. 
In this case, the matrix $R$ takes the $3\times 3$ form, and its one of the explicit forms is given by~\cite{3hdm_ch} 
\begin{align}
R &= 
\begin{pmatrix}
c_{\beta_2} &0 & -s_{\beta_2}  \\
0&1&0 \\
s_{\beta_2} &0 & c_{\beta_2}   \\
\end{pmatrix}
\begin{pmatrix}
c_{\beta_1} & -s_{\beta_1} & 0\\
s_{\beta_1} &  c_{\beta_1} & 0\\
0&0&1 
\end{pmatrix}
=
\begin{pmatrix}
c_{\beta_1} c_{\beta_2} & -s_{\beta_1} c_{\beta_2} & -s_{\beta_2}\\
s_{\beta_1} &  c_{\beta_1} & 0\\
c_{\beta_1} s_{\beta_2}&-s_{\beta_1}s_{\beta_2}&c_{\beta_2} 
\end{pmatrix},~~\text{for}~~N=3,  
\end{align}
where $\tan\beta_1 \equiv v_2/\sqrt{v_1^2+v_3^2}$ and $\tan\beta_2 \equiv v_3/v_1$. 
A similar but different parameterization of the three VEVs has also been given in Ref.~\cite{Diaz-Cruz}. 
Using this notation, each of $\xi_F^a$ factors are expressed as in Table~\ref{ratios}, where 
we assign $\Phi_u = \Phi_2$, and $\Phi_{d,e}=\Phi_1$ (if these are different from $\Phi_u$) in the Type-II, -X and -Y interactions. 
For the Type-Z, we assign $(\Phi_u,\Phi_d,\Phi_e)=(\Phi_2,\Phi_1,\Phi_3)$. 
Unlike the 2HDMs, the second term of Eq.~(\ref{kappaf}) is not simply determined by $\kappa_V^{}$, 
so that to get the prediction of $\kappa_F^{}$, we need to further input $(R_{H})_{21}$ and $(R_{H})_{31}$ under the 
constraint of $(R_H)_{21}^2 + (R_H)_{31}^2 = 1 -\kappa_V^2$. 

We here discuss critical differences of the Higgs boson couplings between in the 2HDMs and in the 3HDMs. 
First, as it is immediately seen, the Type-Z Yukawa interaction is only realized in 3HDMs. 
In other words in 2HDMs, there is at least one pair of $\kappa_F^{}$ and $\kappa_{F'}^{}$ ($F\neq F'$) with $\kappa_F^{}=\kappa_{F'}$ as seen in (\ref{types}), while 
the Type-Z can provide the prediction of $\kappa_u\neq \kappa_d$, $\kappa_d \neq \kappa_e$ and $\kappa_e\neq \kappa_u$. 
Next, in 2HDMs the prediction of $\Delta\kappa_d\neq \Delta\kappa_e$ with $\Delta\kappa_{d,e}>0$ or $\Delta\kappa_{d,e}<0$ is not allowed as explained by the following.  
First of all, $\Delta\kappa_d \neq \Delta\kappa_e$ is realized in the Type-X and Type-Y 2HDM. 
However, in these types, e.g., in the Type-X, $\xi_e=-\tan\beta$ and $\xi_d=\cot\beta$. 
Therefore for a fixed $\sigma$,  the sign of $\Delta\kappa_e$ is opposite to that of $\Delta\kappa_d$ as long as $|\xi_F| >  \sqrt{\Delta\kappa_V^{}/(2-\Delta\kappa_V^{})}$. 
On the other hand, this is not the case for the 3HDMs as it will numerically be shown in the next section.

\section{Deviation in the Higgs boson couplings}

In this section, we show the correlation between $\Delta\kappa_\tau$ and $\Delta\kappa_b$ in the 2HDMs and in the 3HDMs. 
We choose the following parameters as inputs:
\begin{align}
&\Delta\kappa_V^{},~\tan\beta,~\sigma,~~\text{for the 2HDMs}, \\
&\Delta\kappa_V^{},~\tan\beta_1,~\tan\beta_2,~(R_H)_{21},~\sigma',~~\text{for the 3HDMs}, 
\end{align}
where $\sigma'\equiv \text{Sign}[(R_H)_{31}]$. 
In these input parameters for the 3HDM, $|(R_H)_{31}|$ is derived by $\sqrt{1-\kappa_V^2-[(R_H)_{21}]^2}$. 
For all the calculations below, we scan the value of $(R_H)_{21}$ in the range from $-\sqrt{1-\kappa_V^2}$ to $+\sqrt{1-\kappa_V^2}$. 
In order to clarify the possible allowed predictions in the 3HDMs which cannot be explained in the 2HDMs, we take into account the 
one-loop electroweak corrections to the Yukawa couplings in the 2HDMs based on the on-shell renormalization scheme\footnote{In Ref.~\cite{Santos}, improved renormalization schemes 
in the 2HDM have been proposed where gauge dependences in the mixing angles are successfully removed. } according to Refs.~\cite{THDM-loop1,THDM-loop2}. 
For the one-loop calculation, we scan the parameters of the 2HDM by $m_{\Phi}^{} \geq 300$ GeV, $\tan\beta \geq 1$ and $|\lambda_{h\Phi\Phi}^{}|\geq 0$ 
with $\lambda_{h\Phi\Phi}^{}\equiv (m_\Phi^2 - M^2)/v$, 
where $m_{\Phi}^{}$ is the mass of extra Higgs bosons (here we assume that all the extra Higgs bosons are degenerate in mass), and $M$~\cite{KOSY} is a dimensionful parameter which 
is irrelevant to the Higgs VEV (see, e.g., \cite{THDM-loop1} for the detailed explanation about these parameters). 
For the SM inputs, we use the following values~\cite{PDG}
\begin{align}
&m_t = 173.21~\text{GeV},~
m_b=4.66~\text{GeV},~
m_c=1.275~\text{GeV},~
m_\tau=1.77684~\text{GeV},~
m_h=125~\text{GeV}, \notag\\
&\alpha_{\text{em}} =(137.035999074)^{-1},~
m_Z^{}=91.1876~\text{GeV},~
G_F=1.1663787\times 10^{-5}~\text{GeV}^{-2},   \notag\\
&\Delta\alpha_{\text{em}}=0.06635,~\alpha_s=0.1185~\text{GeV}. 
\end{align}
%
We take into account theoretical constraints on the model parameters from the perturbative unitarity~\cite{PU1,PU2,PU3,PU4}, the vacuum stability~\cite{VS}, 
and the triviality with the criterion that the Landau pole does not appear below 3 TeV. 
We note that the amount of QCD corrections to the $hb\bar{b}$ coupling can be a sub percent level in the SM~\cite{Peskin}, which are not included in our calculation.

\begin{figure}[t]
\begin{center}
\includegraphics[width=70mm]{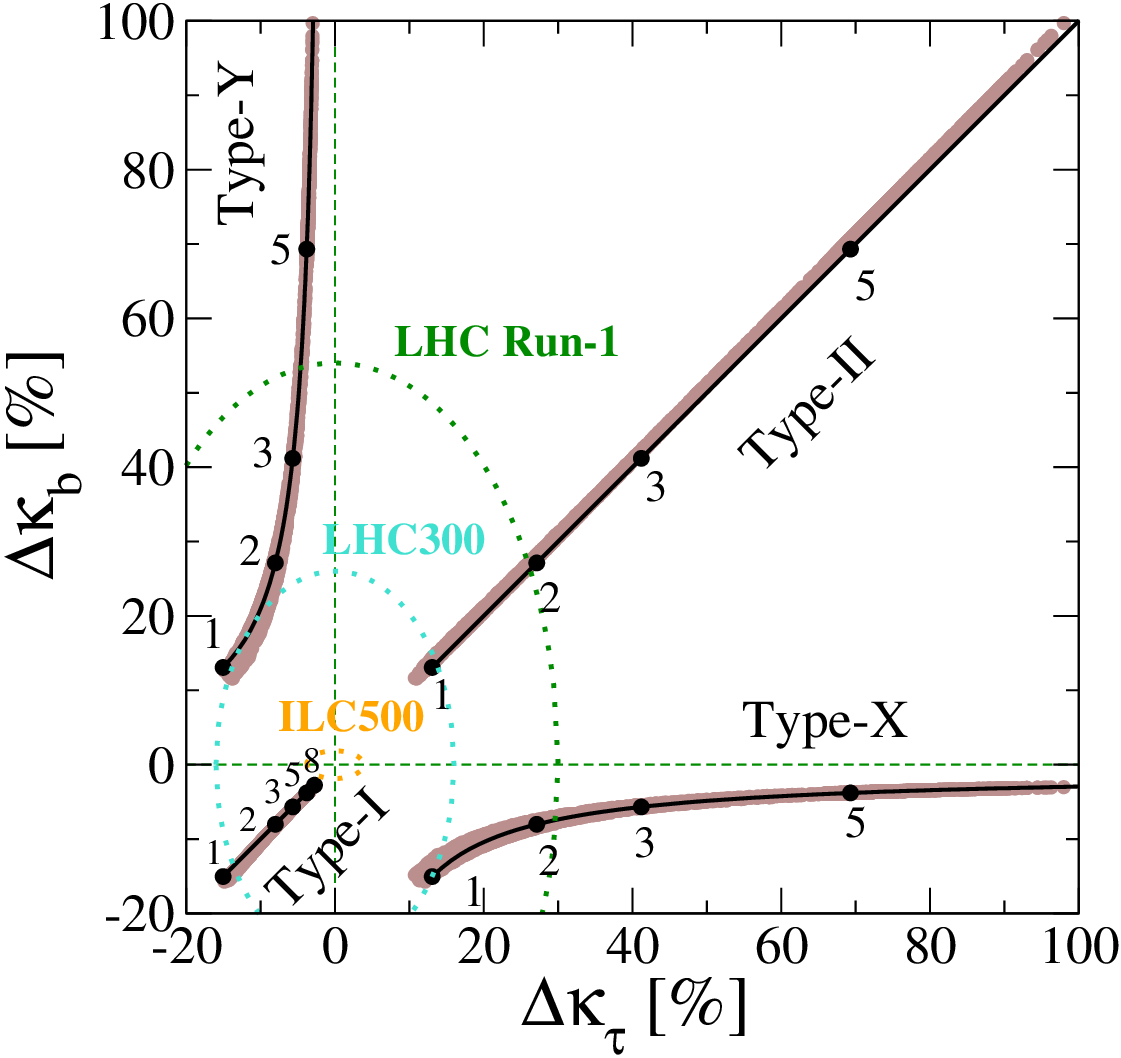}\hspace{5mm}
\includegraphics[width=70mm]{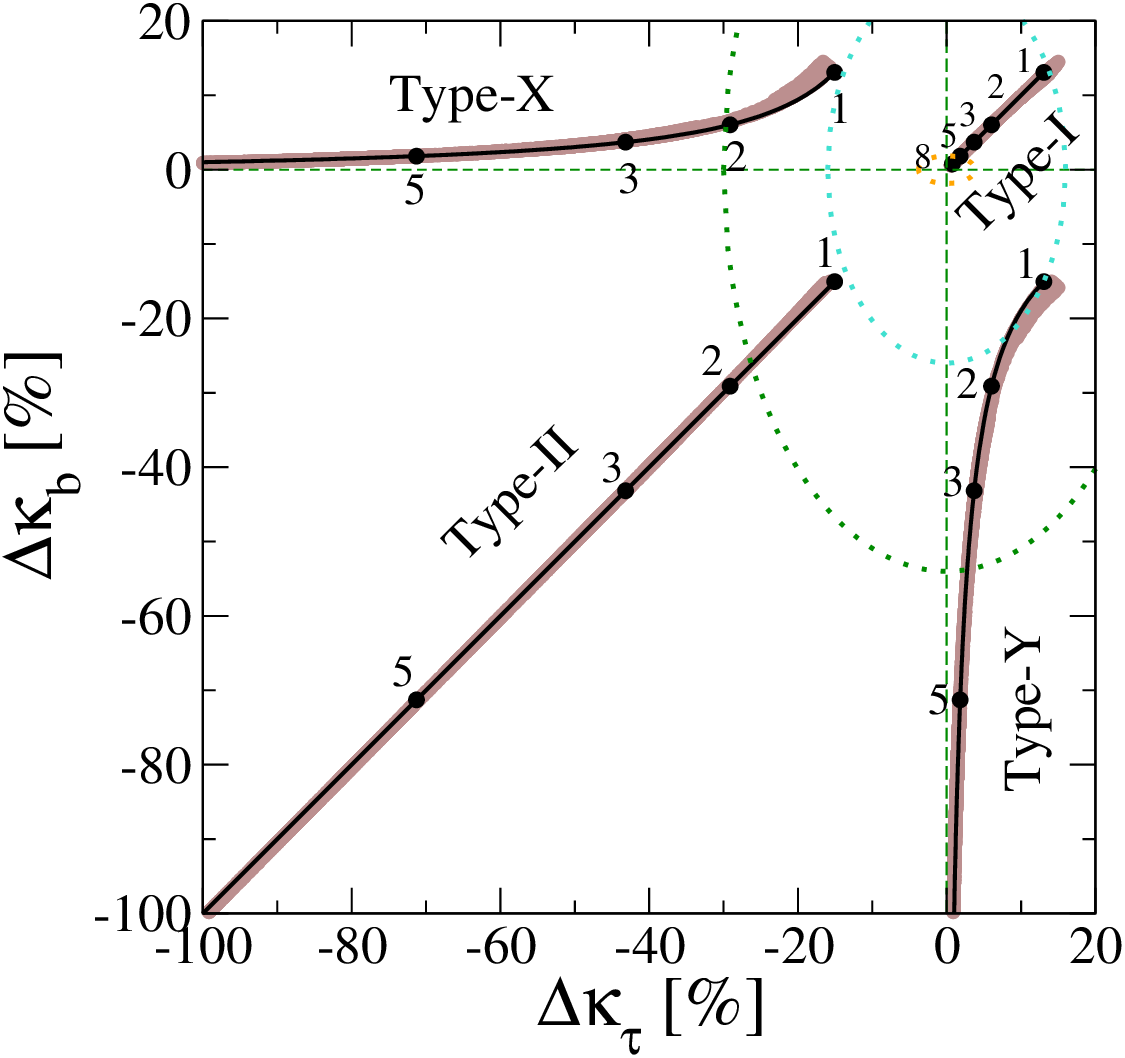}
\caption{Predicitons on the $\Delta\kappa_\tau$ and $\Delta\kappa_b$ plane in the Type-I, -II, -X and -Y 2HDMs in the case of $\Delta\kappa_V^{} = 1\%$. 
The solid curves show the tree level prediction. 
Each dot on these curves denotes the prediction of the fixed value of $\tan\beta$, where its value is written beside the dot. 
The gray shaded regions show the one-loop corrected results. 
The left and right panels respectively show the case for $\sigma =-1$ and $\sigma = +1$. 
The largest (green) dotted ellipse shows the 2$\sigma$ error of the measurement of $\Delta\kappa_\tau$ and 
$\Delta\kappa_b$ from the current LHC data extracted from Ref.~\cite{LHC-combine}, while 
the middle (cyan) and smallest (orange) ones respectively show the expected $2\sigma$ accuracy of the measurement at the LHC with the collision energy of 14 TeV and the integrated luminosity of 300 fb$^{-1}$ and at the ILC with the collision energy of 500 GeV and the integrated luminosity of 500 fb$^{-1}$) extracted from~\cite{HWG}. 
}
\label{fig1}
\end{center}
\end{figure}

First, we show the difference of the prediction of $\Delta\kappa_\tau$ and $\Delta\kappa_b$ in the 2HDMs with four types of Yukawa interaction. 
In Fig.~\ref{fig1}, we show the correlation between $\Delta\kappa_\tau$ and $\Delta\kappa_b$ in the four types of Yukawa interaction in the 2HDM. 
We here take $\Delta\kappa_V^{} (= 1 - (R_H)_{11}) = 1\%$, $\tan\beta \geq 1$ and $\sigma=-1~(+1)$ in the left (right) panel. 
The solid curves show the tree level prediction, while the shaded regions do the one-loop corrected result. 
We can see that the predictions of the four types of Yukawa interaction are well separated in the $\Delta\kappa_\tau$--$\Delta\kappa_b$ plane even taking into account 
the one-loop correction to the Yukawa couplings. 
In this figure, we display the $2\sigma$ error of the measurement of $\Delta\kappa_\tau$ and $\Delta\kappa_b$ 
at the LHC Run-1 experiment\footnote{Here, we extract the $2\sigma$ errors of $\Delta_\tau$ and $\Delta_b$ from Ref.~\cite{LHC-combine}, but we do not use their central values.  } by the 
green dotted ellipse,  
where data from the ATLAS and CMS collaborations are combined~\cite{LHC-combine}. 
In addition, we also show the expected  $2\sigma$ accuracy of the measurement of $\Delta\kappa_\tau$ and $\Delta\kappa_b$
at the LHC with 14 TeV and 300 fb$^{-1}$ (denoted as LHC300) and at the International Linear Collider (ILC) with 500 GeV and 500 fb$^{-1}$ (denoted as ILC500). 
Although the current measurement (LHC Run-1) is not enough accurate to  separate the four types of Yukawa interaction in the configuration with $\Delta\kappa_V^{}=1\%$, 
we may be able to discriminate these types of Yukawa interaction by the future measurements at the LHC300 and the ILC500. 
We note that for a larger (smaller) value of $\Delta\kappa_V^{}$, 
the distance of each prediction in the four types of Yukawa interaction becomes to be more (less) spread. 
As the extreme case, when $\Delta\kappa_V^{}\to 0$ is taken, all the predictions converge at $(\Delta\kappa_\tau,\Delta\kappa_b)\to (0,0)$. 

\begin{figure}[!t]
\begin{center}
\includegraphics[width=70mm]{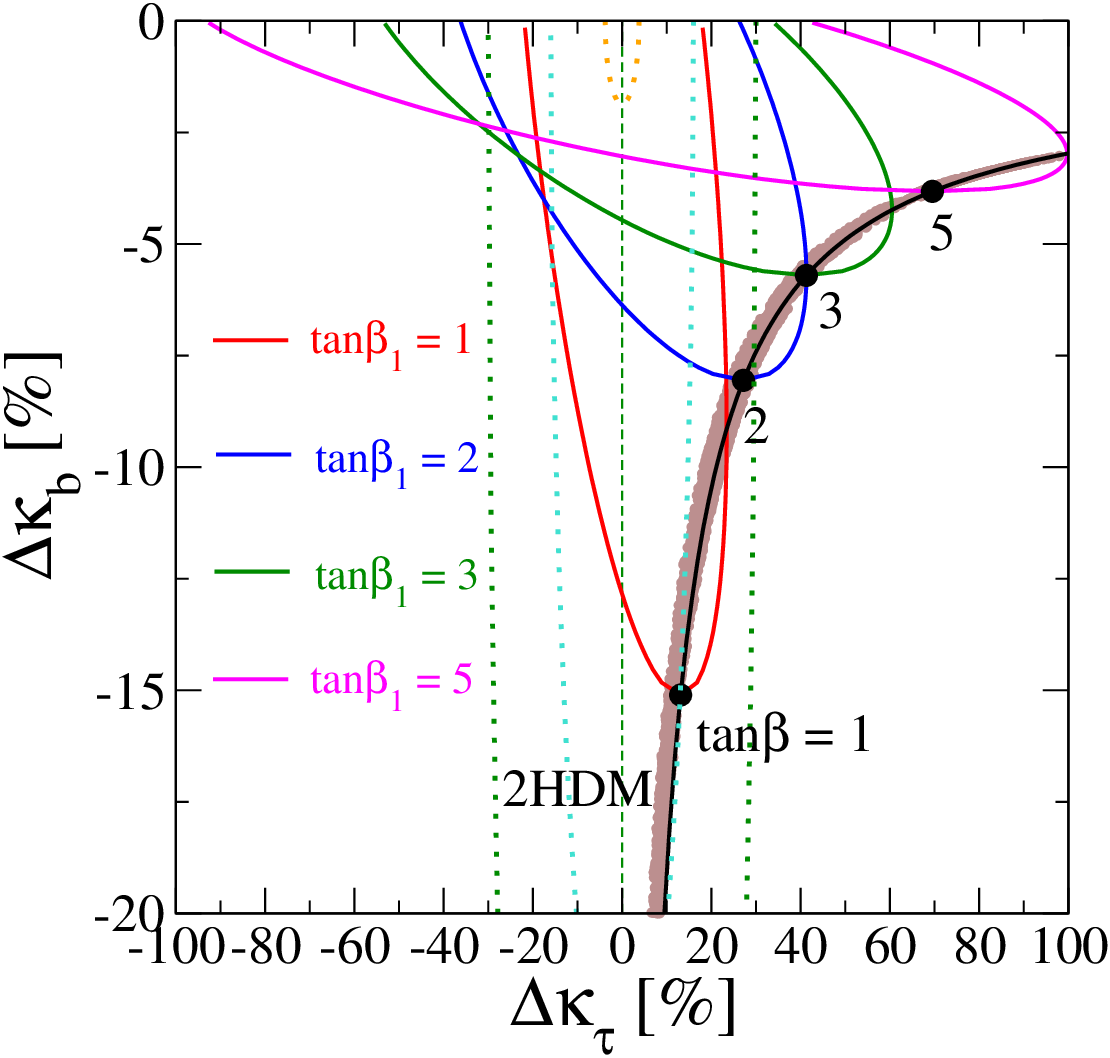}\hspace{5mm}
\includegraphics[width=70mm]{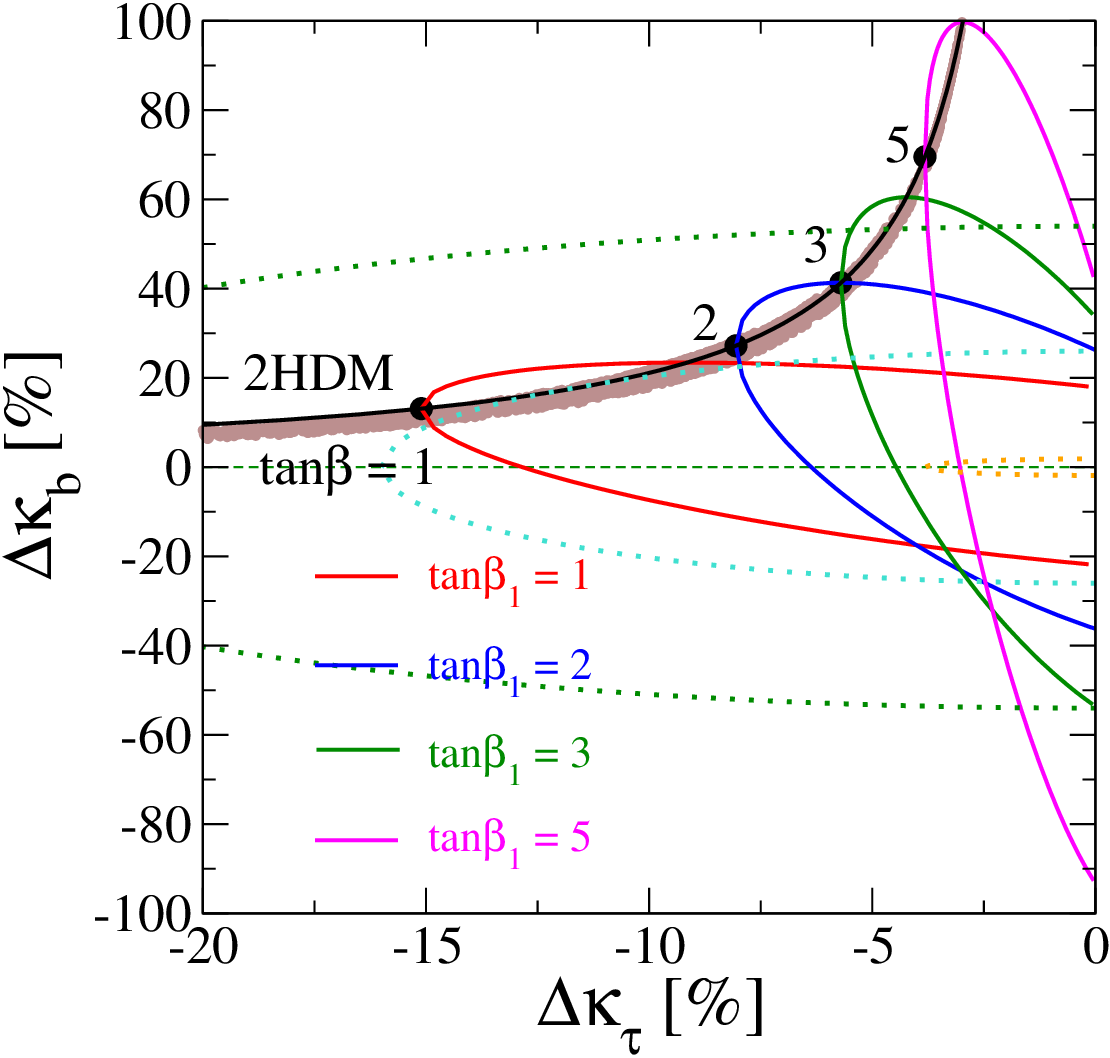}\\  \vspace{10mm}
\includegraphics[width=70mm]{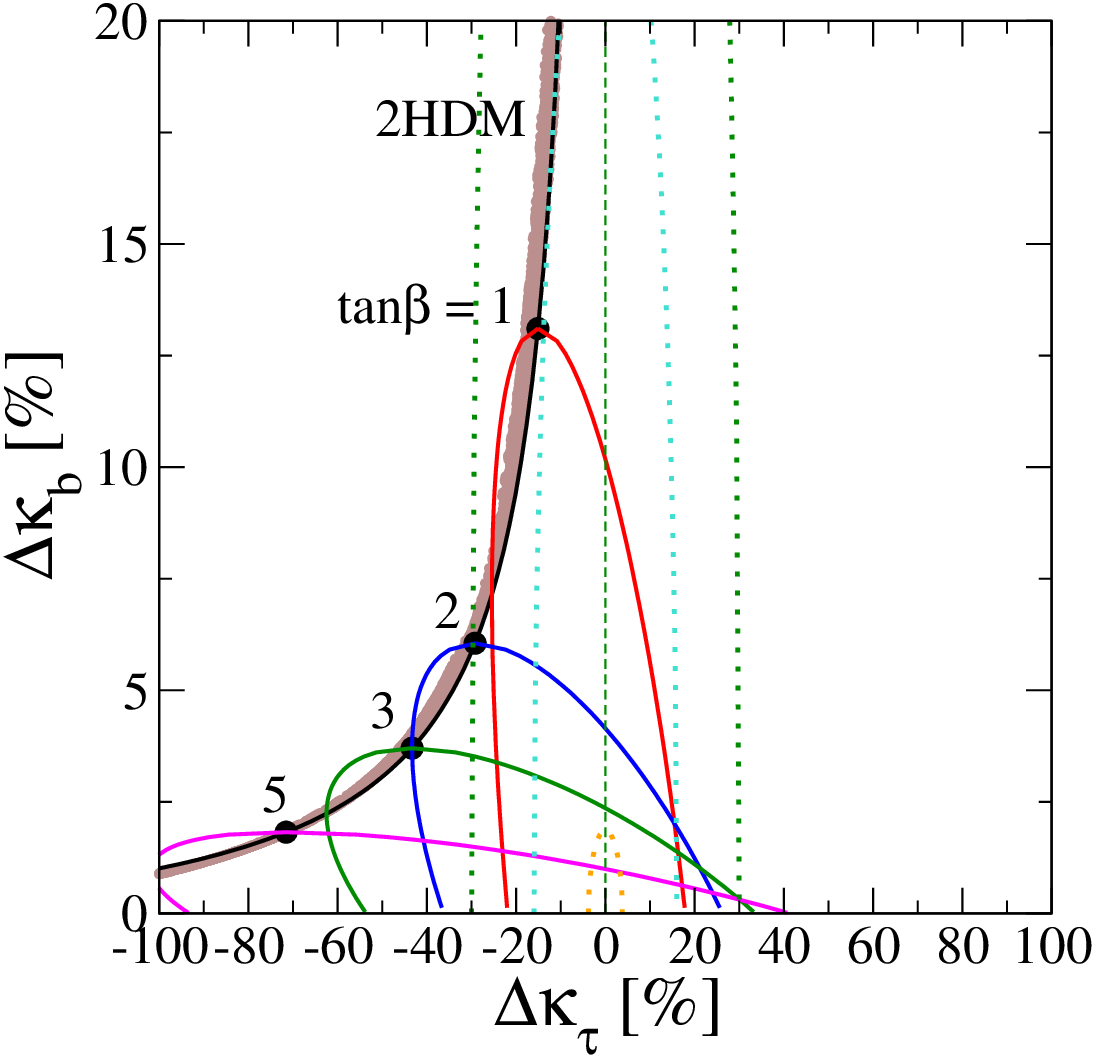}\hspace{5mm}
\includegraphics[width=70mm]{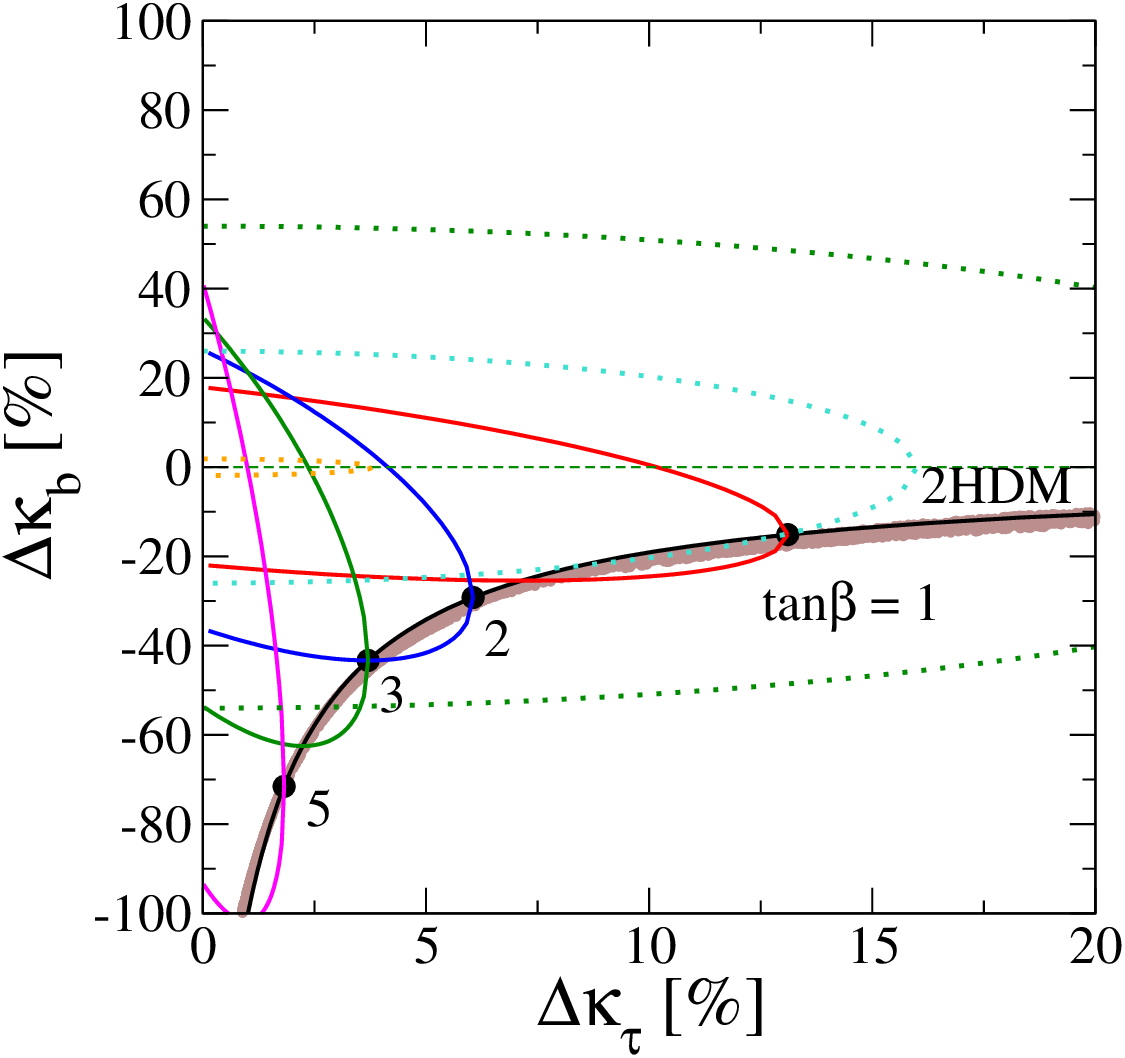}
\caption{Predicitons of $\Delta\kappa_\tau$ and $\Delta\kappa_b$ in the 2HDM and in the 3HDM in the case of $\Delta\kappa_V^{} = 1\%$. 
The left-upper, right-upper, left-lower and right-lower panel respectively shows the case for 
(Type-X, $\Delta\kappa_u < 0$), 
(Type-Y, $\Delta\kappa_u < 0$),
(Type-X, $\Delta\kappa_u > 0$) and 
(Type-Y, $\Delta\kappa_u > 0$). 
Similar to Fig.~\ref{fig1}, 
the black curve shows the tree level prediction, and the gray shaded region does the one-loop corrected result in the 2HDM.  
The red, blue, green and magenta curve respectively shows the tree level prediction in the 3HDM with  
$\tan\beta_1 = 1$, 2, 3 and 5 in the 3HDM. 
For all the results in the 3HDM, we take $\tan\beta_2=1$. 
The description of the three ellipses is the same as those in Fig.~\ref{fig1}. 
}
\label{fig2}
\end{center}
\end{figure}

Next, we compare the prediction of $\Delta\kappa_\tau$ and $\Delta\kappa_b$ in the 2HDM and in the 3HDM. 
Because in the Type-I and Type-II case, tree level predictions in both the 2HDM and the 3HDM are given on the line of $\Delta \kappa_b = \Delta\kappa_\tau$, 
it is difficult to see the difference between these two models. 
We thus compare the prediction in the Type-X and Type-Y Yukawa interaction. 
In Fig.~\ref{fig2}, we show the correlation between $\Delta\kappa_\tau$ and $\Delta\kappa_b$ in the 2HDM and in the 3HDM in the case of $\Delta\kappa_V^{} = 1\%$. 
The results for the Type-X (Type-Y) are given in the left (right) panel, while those in  
the case for $\Delta \kappa_u < 0$ $(\Delta \kappa_u > 0)$ are displayed in the upper (lower) panel. 
Clearly, we can find the region which cannot be drawn by the one-loop corrected prediction in the 2HDM, but can be explained in the 3HDM. 
We note that in the 3HDMs with a larger (smaller) value of $\tan\beta_2$, each of the parabolas becomes that with a larger (small) curvature through the same point given in the 2HDM
denoted by the black dot. 

\begin{figure}[t]
\begin{center}
\includegraphics[width=70mm]{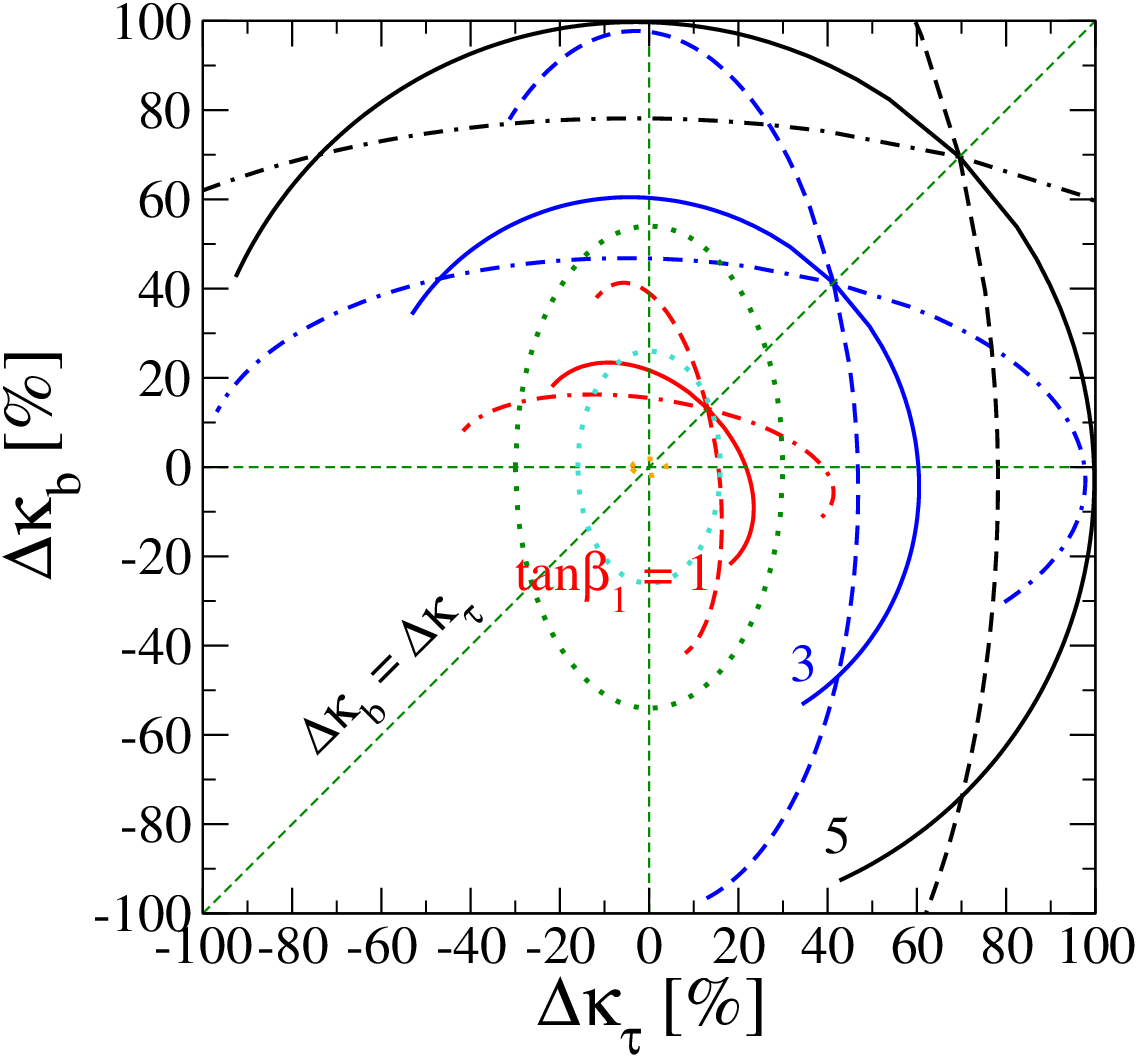}\hspace{5mm}
\includegraphics[width=70mm]{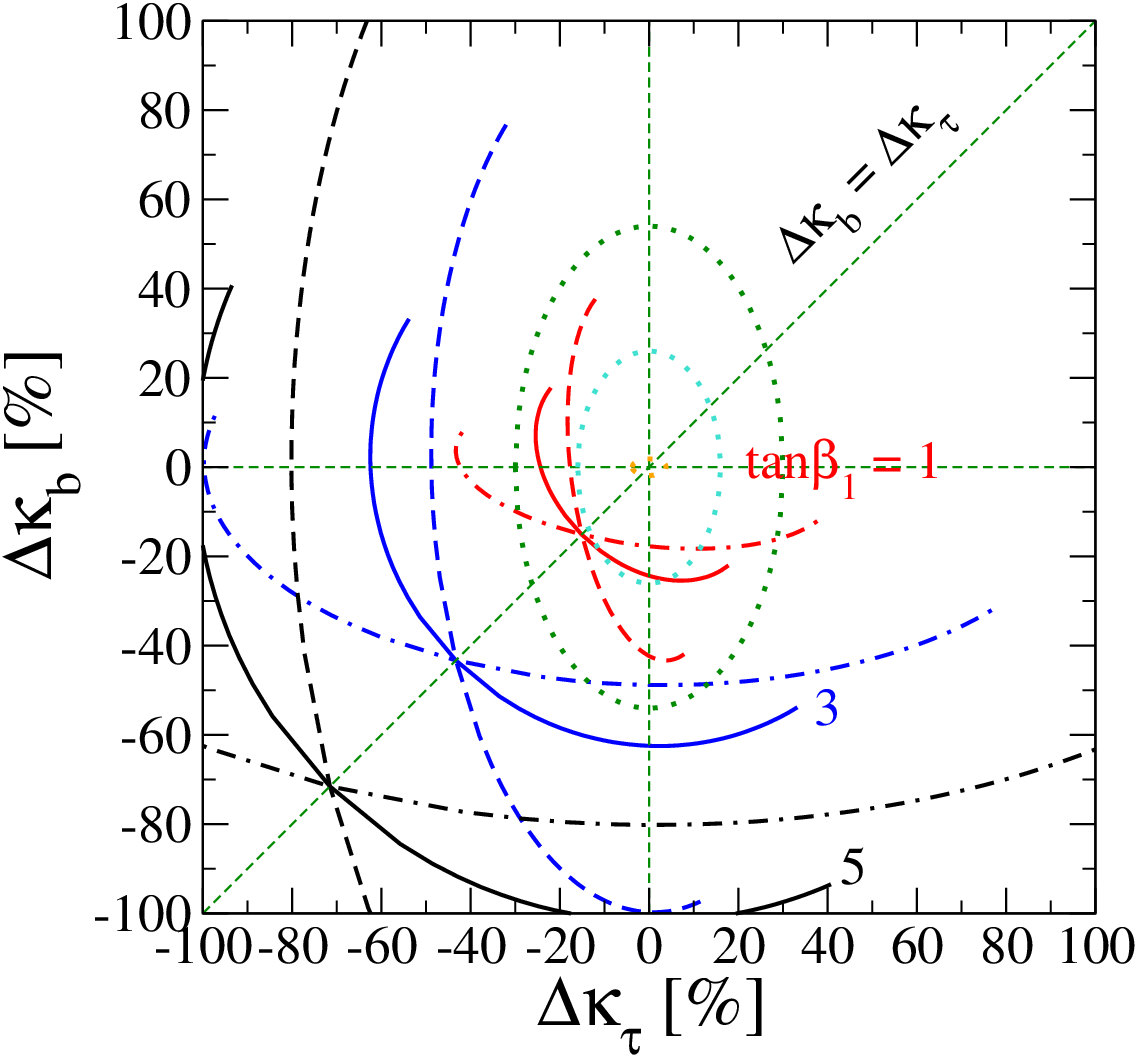}
\caption{Predicitons of $\Delta\kappa_\tau$ and $\Delta\kappa_b$ in the Type-Z 3HDM in the case of $\Delta\kappa_V^{} = 1\%$. 
The left (right) panel shows the case for $\Delta \kappa_u <0$ ($\Delta \kappa_u >0$). 
The value of $\tan\beta_2$ is taken to be 1 (solid curve), 2 (dashed curve) and 1/2 (dotted curve). 
The description of the three ellipses is the same as that in Fig.~\ref{fig1}. }
\label{fig3}
\end{center}
\end{figure}

Finally, we show the prediction in the Type-Z 3HDM on the $\Delta\kappa_\tau$ and $\Delta\kappa_b$ plane. 
The left and right panel show the case for $\Delta \kappa_u <0$ and $\Delta \kappa_u >0$, respectively. 
We can see that the predictions with $\Delta\kappa_{\tau,b}>0$ ($\Delta\kappa_{\tau,b}<0$) 
and $\Delta\kappa_\tau \neq \Delta\kappa_b$ are allowed for the case of $\Delta\kappa_u<0$ ($\Delta\kappa_u>0$), which are not allowed in the 2HDMs at the tree level. 
Even if we take into account the one-loop corrections to the Yukawa coupling, only a few percent level of the difference between $\Delta\kappa_\tau$ and 
$\Delta\kappa_b$  is allowed as we see in Fig.~\ref{fig1}. 
Therefore, if the Higgs boson couplings are measured to be in this region, i.e., $\Delta\kappa_{\tau,b}>0$ ($\Delta\kappa_{\tau,b}<0$) 
and $\Delta\kappa_\tau \neq \Delta\kappa_b$, it can be evidence for the MHDMs with $N\geq 3$. 

\section{Conclusions}

We have studied the SM-like Higgs boson couplings with weak bosons $hVV$ and fermions $hf\bar{f}$ in MHDMs with NFC. 
We have presented the generic expression for the scaling factors of $\kappa_V^{}$ and $\kappa_F^{}$ in the model with $N$ doublet scalar fields. 
As simple concrete examples, we have discussed the 2HDMs and the 3HDMs, and have 
presented the formulae of these scaling factors  
in terms of the ratios of the Higgs VEVs and the matrix elements of the diagonalization matrix for the CP-even scalar states for each type of Yukawa interaction. 
We then have shown the correlation between $\Delta\kappa_\tau$ and $\Delta\kappa_b$ in the 2HDMs and in the 3HDMs under the assumption that 
the $hVV$ couplings deviate from the SM prediction by 1\%. 
It has been clarified that there are predictions in the $\Delta\kappa_\tau$ and $\Delta\kappa_b$ plane in the 3HDMs, 
which cannot be explained in the 2HDMs even when taking into account the one-loop corrections to the Yukawa coupling. 

We would like to mention that such a region can also be explained within a 2HDM if we relax the framework of NFC such as the so-called Type-III 2HDM. 
However, in such a model FCNCs mediated by neutral Higgs bosons are naturally induced, so that  measurements at flavour experiments also become
important to discriminate the models with NFC and those without NFC. 

Finally, we briefly comment on the case with CP-violation in the Higgs sector which is assumed not being occurred in this paper. 
If there is a non-zero physical CP-violating phase in the Higgs potential, CP-even and CP-odd component scalar fields are mixed with each other. 
Through the mixing, the Yukawa coupling for the SM-like Higgs boson contains a term proportional to the $\gamma_5$ matrix, which can modify
various physical quantities such as decay rates and cross sections of the SM-like Higgs boson from the SM prediction even if $\kappa_F =1$ is taken. 
The collider phenomenology with such a CP-violating effect has been studied in Refs.~\cite{Barroso, Fontes, Yagyu-cpv} in the 2HDMs. 
In addition, the method to extract the CP-even component (without $\gamma_5$) and the CP-odd component (with $\gamma_5$) of the $ht\bar{t}$ coupling
has been proposed in Ref.~\cite{Gunion} by measuring the $ht\bar{t}$ cross section weighted by an operator constructed from the (anti-)top quark momentum.

\noindent
\section*{Acknowledgments}
\noindent 
The author is grateful to Shinya Kanemura for fruitful discussions. 
He also thanks to Andrew Akeroyd for careful reading of the manuscript. 
This work was supported by a JSPS postdoctoral fellowships for research abroad.

\end{document}